# Silver Telluride Colloidal Quantum Dot Infrared Photodetectors and Image Sensors

Yongjie Wang[1], Lucheng Peng[1], Julien Schreier[2], Yu Bi[2], Andres Black[2], Aditya Malla[1], Stijn Goossens[2], Gerasimos Konstantatos[1,3*]

[1]ICFO-Insitut de Ciencies Fotoniques, The Barcelona Institute of Science and Technology, Castelldefels, 08860 Barcelona, Spain.

[2] Qurv Technologies S.L., Castelldefels, 08860, Spain

[3]ICREA-Institució Catalana de Recerca i Estudiats Avançats, Lluis Companys 23, 08010 Barcelona, Spain.

*e-mail: Gerasimos.Konstantatos@icfo.eu

**Photodetectors that are sensitive in shortwave infrared (SWIR) range (1 µm - 2 µm) are of significant interest for applications in 3D, night and adverse weather imaging, machine vision and autonomous driving, among others. Currently available technologies in the SWIR rely on costly epitaxial semiconductors that are not monolithically integrated with CMOS electronics. Solution-processed quantum dots can address this challenge by enabling low-cost manufacturing and simple monolithic integration on silicon in a back-end-of-line (BEOL) process. To date, colloidal quantum dot (CQD) materials to access the SWIR are mostly based on lead sulfide (PbS) and mercury telluride (HgTe) compounds, imposing major regulatory concerns and impeding their deployment in consumer electronics due to toxicity concerns. Here we report a new synthesis method for environmentally-friendly silver telluride ($Ag_2Te$) quantum dots and their application in high-performance SWIR photodetectors. The CQD photodetector stack employs materials compliant with the *Restriction of Hazardous Substance* (RoHS) directives and is sensitive in the spectral range from 350 nm - 1600 nm. The room-temperature detectivity is of the order $10^{12}$ Jones, the 3dB bandwidth is in excess of 0.1 MHz and the linear dynamic range is over 118 dB. We also realize a monolithically integrated SWIR imager**

**based on solution processed, heavy-metal-free materials, thus paving the way of this technology to consumer electronics market.**

Shortwave infrared (SWIR) light for sensing and imaging is of paramount importance owing to its unique characteristics: It is eye-safe; It can penetrate through fog, haze, and other atmospheric conditions, enabling imaging under adverse weather for automotive applications, environmental and remote sensing;[1–3] the presence of night glow under night in the SWIR enables passive night vision;[4] visual imaging combined with infrared spectroscopy enables machine vision,[3] bio-imaging,[5–8] food and process quality inspection etc.[9,10] Of particular relevance in the SWIR is the band around 1.35 -1.4 μm that provides an eye-safe window, free of background light under day/night conditions enabling long-range LIDAR and 3D imaging for automotive and Augmented/Virtual Reality (AR/VR) applications.[11]

In view of the huge market potential and the fact that established technologies to address the SWIR are based on highly costly and non-monolithic to silicon III-V technologies, significant efforts over the last 15 years have led to the demonstration of highly performing colloidal quantum dot photodetectors and image sensors.[2,9,12–22] However, to date, the available CQD materials to access the SWIR are based on heavy metals of Pb and Hg that are RoHS regulated elements and currently prohibited for use in consumer electronics.

In this work we sought for a new heavy-metal free colloidal quantum dot semiconductor that would preserve the favorable properties of its heavy-metal counterparts in terms of broad bandgap tuneability and facile solution processing. We identified silver telluride quantum dots ($Ag_2Te$ CQDs), as promising contender and after a thorough literature review we realized that while $Ag_2Te$ CQDs have been synthesized and applied in bio-imaging applications as fluorophores in the near infrared and as a template to form efficient p-type doped HgTe CQD hole transporting layer in HgTe QDs photodiodes,[2,15,23–26] its demonstration as highly performant absorbers with tuneable and strong optical absorption akin to other colloidal quantum dot materials has remained elusive. So far, reported silver chalcogenide CQDs operate in photo-conductor mode featuring high dark current and low operating speed.[24,27–32] The only reported performance of $Ag_2Te$ QDs based photodiode showed an external quantum efficiency less than 0.5% in the near infrared region.[33]

We posited that the origin of the challenge to achieve wide size tunability and thus broadband tunable optical properties, lies in the limited choice of tellurium precursors.[26,34] Up to now, the most widely used tellurium precursor has been tri-alkyl-phosphine-tellurium ($R_3P = Te$), while phosphine has been suspected to adversely affect surface passivation, thereby the photoluminescence quantum yield.[26] At the same time, phosphine-based ligands are calculated to bind more strongly to Ag on the surface of QDs than thiols, [26,35] and are probably not an optimal ligand choice for CQDs that would need to undergo a ligand exchange process to render the films electrically conductive.

**Phosphine-free $Ag_2Te$ CQDs.**

Various methods and precursors have been used to synthesize $Ag_2Te$ QDs under different reaction conditions.[23,25,36,37] However, achieving size tunability in these QDs has been challenging. Recently, large-sized $Ag_2Te$ QDs with a photoluminescence peak exceeding 2000 nm were obtained using phosphine-induced ripening.[26] Nevertheless, obtaining a decent excitonic absorption peak above 1500 nm remains elusive, indicating insufficient control over size uniformity.[38] This limitation leads to energetic disorder and hinders charge transport, thereby compromising the performance of optoelectronic devices.[39–41] Therefore, there is an urgent need to develop a synthetic method capable of producing monodisperse $Ag_2Te$ QDs with distinct excitonic peaks, which can be exploited in promising optoelectronic applications.

Herein, we introduce a phosphine-free tellurium precursor that led us obtain $Ag_2Te$ QDs with well-controlled size distribution and distinct excitonic peaks over a very broad range (800 nm to 2000 nm). Silver-oleylamine and tellurium-thiol complexes were used as precursors for hot injection, wherein tellurium oxide was reduced by thiols[42] and further reacted with silver precursors (Supplementary Note 1). Thanks to the absence of phosphine, $Ag_2Te$ QDs are stable in crude solution for hours under heating without showing Ostwald ripening, enabling us to easily tune sizes with controlled precursor supply. As shown in Fig. 1a, the absorption spectrum of $Ag_2Te$ QDs showed clear excitonic peaks, whose positions were controlled by reaction temperatures or continuous injection method (See Experimental Methods). The $Ag_2Te$ QDs showed compelling colloidal stability upon storing in air for two weeks (Supplementary Fig. 1). Intense photoluminescence was also observed under near infrared light excitation (Supplementary Fig. 2). More importantly, we report here for the first time, $Ag_2Te$ QDs with distinct excitonic peaks (peak-valley ratio > 1.2) over 1500 nm, which has been a formidable obstacle for phosphine-based methods.[26,34]

X-ray photoelectron spectroscopy was first employed to examine the material composition and stoichiometry as function of QD size (Supplementary Fig. 3). The Te to Ag ratio increases with the bandgap decreasing (size increasing), matching well with the calculated ratio of a cation-rich surface configuration (Supplementary Fig. 3c&d). Transmission electron microscopy (TEM) images (Fig. 1b-k) confirmed the controllable and tunable QD sizing from less than 2.9 nm to over 6.6 nm, with regular spherical shapes. High resolution TEM (Fig. 1l) showed high quality single-crystal characteristics with an evident lattice spacing of 0.23 nm, with terminate facets being $(\bar{2}04)$, $(211)$ and $(210)$ (Supplementary Fig. 4). XRD patterns of various sizes $Ag_2Te$ QDs further confirmed the *Hessite* crystal phase, matching well with the high resolution TEM images (Supplementary Fig. 5). Thanks to the wide size and consequent bandgap tunability, we were able to extract the sizing curve from TEM size distribution and absorption spectra. The second derivatives of absorption spectra were used to extract the multiple excitonic peaks (Supplementary Fig. 6). The sizing curve was further fitted into equation:[43]

$$E_g = \frac{1}{2}\left[E_0 + \sqrt{{E_0}^2 + 8E_0\alpha\pi^2 \frac{R_y a_0}{\varepsilon_\infty d_0} \frac{e^{-\frac{d}{d_0}}}{(1+e^{-\frac{d}{d_0}})^2}}\right]$$

where $E_g$ is the size dependent bandgap, $E_0$ is the bulk bandgap, $R_y$ is the Rydberg energy (13.606 eV), $a_0$ is he Bohr radius of hydrogen (0.053 nm), $\alpha$ is a prefactor accounting the size dispersion (~0.7), $\varepsilon_\infty$ is the dielectric constant (here, ~16),[44] $d_0$ is the exciton Bohr diameter, $d$ is the QD diameter. By fitting the first excitonic peak with QD sizes (Fig. 1m), the exciton Bohr diameter was estimated to be ~10 nm, and the bulk bandgap $E_0$ was ~0.53 eV. We note that there exist various reports of the bulk bandgap of $Ag_2Te$ with different values, which may stem from the various crystal phases and non-stoichiometry effects of this semiconductor compound.[45–47] We believe further investigations, including theoretical calculations, would provide more insights.

**$Ag_2Te$ CQDs SWIR photodiode.**

Having obtained $Ag_2Te$ QDs with good monodispersity, we aimed to implement them into a low-temperature, solution-processed, non-toxic SWIR photodiode device that is compatible with complementary metal–oxide–semiconductor (CMOS) readout integrated circuits (ROICs). $Ag_2Te$ QDs photodiodes were initially fabricated with a configuration of ITO/$SnO_2$/$Ag_2Te$ QDs/Au (Fig. 2a). However, devices showed poor rectification with high reverse bias dark current (Fig. 2c). Based on band positions from ultraviolet photoelectron spectroscopy (UPS, Supplementary Fig. 7), we suspected that the poor device performance originates from severe interface recombination between $SnO_2$ and $Ag_2Te$ QDs (Fig. 2b) and therefore sought to introduce a buffer layer that would act as an intermediate electron acceptor offering lower carrier density and better band alignment than $SnO_2$. We found that inserting a buffer layer of $AgBiS_2$ nanocrystals between $SnO_2$ and $Ag_2Te$ QDs improves drastically the device diode quality, leading to a reduced reverse bias dark current density down to ~ 2 μA/cm$^2$ @ -0.1V and ~6 μA/cm$^2$ @ -0.5 V (Fig. 2c and Supplementary Fig. 8). Further diode analysis showed that devices without $AgBiS_2$ NC interface layer have severe non-Ohmic leakage from interface recombination current and a very low dark current activation energy (Supplementary Note 2). In addition to dark current reduction, the external quantum efficiency (EQE) showed substantial improvement with $AgBiS_2$ NCs buffer layer, together with a red-shift of EQE peak position, primarily due to optical Fabry-Perot cavity effect (Supplementary Fig. 9). The optimal photodiodes reached an EQE up to ~20% @1400 nm under 0 V bias (Fig. 2d) and 30% @-0.1 V (Supplementary Fig. 10), while the devices without $AgBiS_2$ NCs showed peak EQE of only ~5%. In comparison, devices with phosphine based methods showed poor rectification and very low photo-responsivity (Supplementary Note 3), similar to previous reports,[33] further suggesting the superiority of our phosphine-free synthetic method.

Considering its importance, device stability was assessed in terms of dark *JV* curves and photo-responsivity. Un-encapsulated devices showed decent air stability maintaining original performance after 25 days' air storage and 180 hours' continuous operation under SWIR light (Supplementary Note 4). The linear dynamic range (LDR) of optimal $Ag_2Te$ QD photodiode was further evaluated by varying the incident light intensity. As shown in Fig. 3a, the photocurrent maintained linearity with incident light power over a broad range and LDR is estimated to be over 118 dB, which compares favorably over PbS or HgTe QDs based photodiodes.[17,48] Frequency dependent response was measured under zero bias, showing a -3dB bandwidth up to 110 kHz (Fig. 3b), which is in accordance with the fast rise and decay times of 1.3 μs and 3.3 μs under modulated light excitation (Fig. 3b inset). Frequency dependent noise spectral density was further measured with fast Fourier Transform (FFT) current transients for low frequency and a lock-in amplifier for high frequency range (Supplementary

Fig. 11). The obtained noise spectrum showed a 1/f noise dominating at low frequency and a generation-recombination noise at high frequency. The device reaches a noise floor of ~$10^{-14}$ A $Hz^{-0.5}$ at frequencies over 20 kHz, which is comparable with many PbS QDs based photodiodes.[17,49] The measured room-temperature specific detectivity ($D^*$) was thus calculated according to:

$$D^* = \frac{R\sqrt{A\Delta f}}{i_n}$$

where $A$ is the device area, $\Delta f$ is the noise bandwidth (1 Hz), $i_n$ being the noise spectral density. The frequency dependent $D^*$ reached over $10^{10}$ *Jones* at 1 Hz and a maximum of ~ $3 \times 10^{12}$ *Jones* at 20 kHz (Supplementary Fig. 12). Moreover, the photodiode showed a $D^*$ over $10^{12}$ *Jones* across a very broad spectral range covering from 350 nm to 1500 nm (Fig. 3c). These results support the potential of $Ag_2Te$ CQD photodetectors in broadband SWIR imaging as well as machine vision applications.

Furthermore, with the size tunability of $Ag_2Te$ QDs developed here, we further extended the spectral peak to 1500 nm by employing larger QDs. The devices with 1520 nm absorption peak QDs showed responsivity of over 0.1 A/W at 1500 nm, -3 dB bandwidth of 220 kHz, $D^*$ reaching $8.8 \times 10^{11}$ *Jones* at 1500 nm (Supplementary Fig. 13).

To the best of our knowledge, the photodiode devices reported here realized, for the first time, solution-processed, non-toxic shortwave infrared photodiodes with figures of merit on par with other heavy-metal containing counterparts (Fig. 3d). These results further support the fact that $Ag_2Te$ QDs emerges as a promising RoHS-compliant material for low-cost, high performance SWIR photodetector applications.

**$Ag_2Te$ CQDs SWIR imager.**

Having developed a heavy-metal-free CQD photodetector, we were then motivated to demonstrate its proof-of-concept potential in a SWIR image sensor use case. Prior to direct integration onto a read-out-integrated circuit (ROIC) we sought to optimize and evaluate its performance in a top-illuminated configuration on a mimic ITO-coated glass substrate (Fig. 4a) using patterned gold as bottom electrodes and 10 nm thin Au as a top semi-transparent electrode. The top-illuminated photodiode exhibited rectification characteristics under dark, albeit inferior to that of the bottom-illuminated devices. Due to the use of ultra-thin Au as a top electrode, forward bias larger than 0.1V led to device damage, preventing us to apply larger forward bias. The detector reaches an EQE ~9% at 1200 nm in short-circuit conditions, further increasing with reverse bias (Fig. 4b). Sputtered ITO was also tried to replace Au as top electrodes, however the devices showed very low responsivity (Supplementary Fig. 14), presumably due to the damage on $Ag_2Te$ QDs upon ITO sputtering.[17]

We thus employed Au-top-electrode-based stack for an image sensor by integrating it on a read-out-integrated circuit (ROIC) to form a focal plane array (FPA) (see methods for details). Fig. 4d illustrates a schematic of the $Ag_2Te$ CQD image sensor. Fig 4e illustrates the fabricated die in the electronic printed circuit board enclosure and Fig. 4f shows a zoomed-in photo of the die comprising the QD stack atop the ROIC. The performance of our $Ag_2Te$ CQD imager was briefly evaluated with printed ISO-12233 test chart and USAF Resolution Target (Supplementary Note 5). Fig. 4g-i illustrate a few examples of images taken using this imager under SWIR light exposure (detecting range 900 nm – ~1600 nm) along with photos taken with a standard mobile phone silicon-based camera. The figures show the transmission of silicon wafers under SWIR light as well as the possibility to visualize the content of plastic bottles that are opaque in the visible. The image sensor performance is envisioned to be further improved building upon recent progress on high quality image sensors reported based on Pb- and Hg-based CQDs[17,20–22].

## Conclusion.

In conclusion, we have first developed a novel synthetic method for size-tunable, phosphine-free $Ag_2Te$ quantum dots. Photodiodes based on these $Ag_2Te$ QDs further demonstrate compelling performance in shortwave infrared region with wide linear dynamic range over 118 dB, -3 dB bandwidth over 110 kHz, and a high room-temperature specific detectivity over $3\times10^{12}$ *Jones*. We further employed heterogeneous monolithic integration of $Ag_2Te$ CQD photodiodes onto Silicon ROICs to demonstrate for the first time a non-toxic, solution-processed SWIR imager operating at room temperature. The results reported here not only circumvent prior challenges in the synthesis of size-tunable, phosphine-free $Ag_2Te$ QDs, but also unleash their potential as a novel RoHS-compliant infrared size-tuneable optoelectronic material. Finally, our work paves the way for the introduction of SWIR colloidal quantum dot technology in consumer electronics markets.

## Acknowledgements.

G.K. acknowledges financial support from the European Research Council (ERC) under the European Union's Horizon 2020 research and innovation programme (grant agreement No 101002306), the Fundació Joan Ribas Araquistain (FJRA), the Fundació Privada Cellex, the program CERCA, and 'Severo Ochoa' Centre of Excellence CEX2019-000910-S funded by the Spanish State Research Agency. Y.W. acknowledges support from the European Union's Horizon 2020 research and innovation programme under the Marie Skłodowska-Curie grant agreement no. 754558. L.P. acknowledges support from the Horizon Europe Framework Programme, the Marie Skłodowska-Curie Postdoctoral Fellowships (grant agreement no.101052595). S.G. acknowledges support from the Horizon – EIC action under the grant agreement no. 101113088 – QSTACK. We would like to thank Yurong Ren, Dr. Debranjan

Mandal for their help during metal electrodes deposition and Dr. Timofiy Khodkov for supplying the diced chips for the imager.

**Author contributions.**

G.K. supervised and directed the study. Y.W. and G.K. conceived the idea, designed this study and co-wrote the manuscript, with feedback from the co-authors. Y.W. synthesized the materials, performed the material characterization, fabricated and characterized the devices, and analyzed the data. L.P. contributed to the development and synthesis of the material. Y.W. fabricated the image sensor with the help of A.B. and Y.B. Y.B. contributed to the design of experiment for the integration of QD film onto the ROIC. J.S. performed the characterization of the image sensor and took the images. Y.W. and J.S. did the performance analysis of the imagers. A. M. contributed to the device fabrication. S.G. contributed in the supervision and design of experiments for the image sensor demonstration.

**Competing interests.**

G.K and Y.W. have filed patent application. G.K. serves as co-founder, share-holder and scientific advisor at Qurv. J.S. and A.B. are employees of Qurv. S.G. is shareholder and employee of Qurv. The remaining authors declare no competing interests.

## FIGURES

**Fig. 1 | Size-tunable $Ag_2Te$ QDs.** **a**, Absorption spectra of various-sized $Ag_2Te$ QDs. Transmission electron microscope (TEM) images of $Ag_2Te$ QDs with absorption peak at **b**, 1310 nm, **d**, 1430 nm, **f**, 1520 nm, **h**, 1750 nm, **j**, 1940 nm, and **c**, **e**, **g**, **i**, **k** their size distribution histograms (*n* = (**c**) 90, (**e**) 117, (**g**) 179, (**i**) 123, (**k**) 288), respectively. Scale bars are 20 nm in **b**, **d**, **f**, **h**, and **j**. TEM sample grids, except the 1940 nm one, were immersed in methanol to remove organic residues for better resolution. **l**, High resolution TEM image of a single QD, showing single crystalline features. Scale bars are 2 nm in both **l** and the inset. **m**, Sizing curve of $Ag_2Te$ QDs from absorption peak positions and TEM size distribution. Dash line indicates the fitting line of first exciton peaks. Error bars indicates the standard deviation of sizes (from **c**, **e**, **g**, **i**, **k**) and full width half maximum (FWHM) of absorption peaks.

**Fig. 2 | $Ag_2Te$ QD Shortwave infrared photodiode.** **a**, Schematics of $Ag_2Te$ QD photodiodes, with and without $AgBiS_2$ NCs buffer layer. **b**, Band diagram of photodiode device. **c**, Dark current density-voltage (*J-V*) curves of photodiodes with and without buffer layer. **d**, External quantum efficiency (EQE) spectrum and responsivity (*R*) of $Ag_2Te$ QD photodiode with and without $AgBiS_2$ NCs buffer layer.

**Fig. 3 | Performance of $Ag_2Te$ QD photodiode.** **a**, Linear dynamic range (LDR) of $Ag_2Te$ QD photodiode under 1310 nm pulse light illumination with power densities from ~10 $nW/cm^2$ to ~10$mW/cm^2$. **b**. Response bandwidth of $Ag_2Te$ QD photodiode. **c**. Specific detectivity spectrum at 20 kHz. **d**. Wavelength dependent specific detectivity of shortwave infrared photodiodes reported in literature.[17,48–55]

**Fig. 4 | Top-illuminated $Ag_2Te$ QD photodetector and imager.** **a**, Schematic of top-illuminated $Ag_2Te$ QD photodetector. **b.** Current density-voltage curves under dark and 0.7 $mW/cm^2$ 1200 nm light illumination**. c.** EQE spectrum of top-illuminated device. **d**. Schematic of ROIC integrated $Ag_2Te$ QDs SWIR imager. **e**. Photo of the imager and **f** zoom in of the ROIC die. **g**. Photos of a silicon wafer on ICFO logo with smartphone visible camera and $Ag_2Te$ QD SWIR camera. **h**. photos of hand holding silicon wafer with visible camera and $Ag_2Te$ QD SWIR camera. **i**. Photos of a dark bottle with liquid using smartphone visible camera and $Ag_2Te$ QD SWIR camera.

## Methods

**Chemicals.** Reagents were purchased from Sigma-Aldrich, except Tellurim (IV) oxide, which is from Thermo Fisher.

**Synthesis of $AgBiS_2$ nanocrystals.** The synthesis of $AgBiS_2$ nanocrystals followed previous reports.[56,57] Briefly, silver acetate (3.2 mmol) and bismuth acetate (4 mmol) were loaded into a flask together with 24 mL oleic acid and 15 mL 1-octadecene. The flask was then pumped under 100 °C to reach a pressure < 0.2 mbar. 4 mmol hexamethyldisilathiane with 5 ml pre-dried 1-octadecene was quickly injected into the flask after switching the flask into Argon atmosphere. The heating was then stopped and the reaction was quenched by water bath. The nanocrystals were purified with acetone for three times and finally dissolved in anhydrous toluene for further use.

**Synthesis of $Ag_2Te$ quantum dots.** $Ag_2Te$ QDs are synthesized with Schlenk line technique from silver iodide-oleylamine (AgI-OLAm) and Tellurium Oxide-dodecanethiol ($TeO_2$-DDT) precursors. Typically, 30 mL oleylamine (OLAm) and 15 mL 1-octadecene (ODE) were loaded into a three-neck flask and pumped at 100°C to remove oxygen and moisture. The flask was then switched into argon atmosphere.

*AgI-OLAm precursor*: silver precursor was prepared by dissolving 10 mmol silver iodide into 10 mL pre-dried OLAm at room temperature under inert atmosphere. Slight heating would facilitate the dissolution process.

*$TeO_2$-DDT precursor*: tellurium precursor was prepared by dissolving 2.5 mmol tellurium oxide ($TeO_2$) into 10 mL dodecanethiol (DDT) under 100 °C heating and inert atmosphere.

The flask containing OLAm and ODE was then heated up to desired temperatures for different size of QDs. Once the temperature reached set point. 3 mL AgI-OLAm precursor was added into the flask. After ~5 min, 5 mL $TeO_2$-DDT precursor was injected into the flask swiftly. The precursor changed into dark black color immediately, indicating the nucleation of $Ag_2Te$ QDs. After 15 min, the heating was stopped and left cooling down naturally. The crude solution was then transferred into inert gas filled glovebox for purification. Methanol was added into the crude solution to precipitate the QDs, followed by 6000 revolutions per minute for 5 min centrifuge. The solid pellet was then fully dispersed into ~30 mL toluene. 3 mL AgI-OLAm precursor was then added into the QD solution, followed by 6000 revolutions per minute centrifuge to remove any precipitates. ~15 mL methanol was added into the supernatant, followed by centrifuge again. The precipitate was discarded and the supernatant, where the QDs are, was kept. ~30 mL methanol was further added into the solution to fully precipitate the QDs, followed by centrifuge. Finally, the solid pellet was collected and the supernatant was discarded. The QD solid was dissolved into anhydrous toluene at various concentration for further characterization and device fabrication.

For the synthesis of large (> 1600 nm) $Ag_2Te$ QDs, continuous injection and growth method was used. Briefly, after injecting 3 mL AgI-OLAm and 5 mL $TeO_2$-DDT precursor at 140 °C, the flask was maintained at this temperature. 1 mL AgI-OLAm and 1.8 mL $TeO_2$-DDT precursor was added dropwise and sequentially with a syringe pump at ~0.05 mL/min. Multiple cycles were executed according to desired sizes. The purification followed the same procedure as previously described for smaller dots.

**Characterization of $Ag_2Te$ QDs.** UV-vis absorption measurements were performed with a Cary 5000 UV-Vis-NIR spectroscope in solution. For photoluminescence measurements, four channel Thorlabs laser was used as excitation light and a Kymera 328i spectrograph (Oxford Instruments, Andor) was used as the detector ( < 1600 nm). The XRD data were collected using a Rigaku SmartLab powder diffractometer in the Bragg–Brentano geometry with Cu Kα radiation on drop-casted powder samples. TEM was performed at the Scientific and Technological Centres of the University of Barcelona. The TEM images were obtained using a JEOL 2100 microscope operating at an accelerating voltage of 200 kV. TEM samples were prepared by dropping diluted QD solution on ultrathin carbon grids, followed by immersing into pure methanol to remove organics for higher resolution. The XPS/UPS measurements were performed with a SPECS PHOIBOS 150 hemispherical analyser under ultrahigh-vacuum conditions ($10^{-10}$ mbar) at the Institut Català de Nanociència i Nanotecnologia.

**Device fabrication.** All the device fabrication steps were carried out in ambient air. ITO-covered glass substrates (Universität Stuttgart, Institut für Großflächige Mikroelektronik) were cleaned by sonication in soapy water, acetone and isopropanol for 20 min each and dried with nitrogen, followed with 0.5 h ultraviolet/ozone treatment. $SnO_2$ electron transport layer was then deposited with an ALD technique at 150 °C to achieve a thickness of ~20 nm. Two layers of $AgBiS_2$ NCs were further spin-coated from 20 mg $ml^{-1}$ toluene solution via the layer-by-layer method. For each $AgBiS_2$ NC layer, one drop of $AgBiS_2$ NC solution was spin coated onto $SnO_2$/ITO substrates during spinning (2,000 r.p.m.). Then, 1,2-ethanedithiol (EDT)/acetonitrile (ACN) (0.1% v/v) solution was applied to the NC film for 20 s, followed by two rinse–spin steps with acetonitrile and once with toluene. Five layers (~100 nm) of $Ag_2Te$ QDs were further deposited on top of $AgBiS_2$ NCs. For each $Ag_2Te$ QD layer, 50 mg $mL^{-1}$ QD solution was spin-coated at 3000 r.p.m. and ligand-exchanged with EDT/ACN (0.1%) for 30 s, followed by rinse twice with ACN and once with toluene. The films were completed by annealing in air for 10 min at 100 °C. Finally, a Kurt J. Lesker NANO 36 system was used to deposit 100 nm Au as top electrode with a device area of 3.1 $mm^2$. For top-illuminated devices, 5 nm Ti and 50 nm Au were deposited onto glass substrates using shade masks to have 1 mm wide stripes. 25 nm $SnO_2$ was deposited by ALD technique, followed by 40 nm $AgBiS_2$ nanocrystal layer and 60 nm $Ag_2Te$ QD layer. 10 nm thin Au was further deposited as top semitransparent electrode with 1 mm stripe shade mask to define a device area as 1 $mm^2$. For imager, we used a global shutter ROIC with a resolution of 640 x 480 pixels that is read-out in an integrate then read (ITR) fashion. The ROIC features two gain modes and programmable detector bias. The in-pixel amplifier is a capacitive trans-impedance amplifier (CTIA). Kapton tape was used to isolate peripheral electrode pads from the detector stack that comprised: 25 nm ALD-$SnO_2$, 40 nm $AgBiS_2$ NCs, 60 nm $Ag_2Te$ QDs and 10 nm thin Au as top electrode.

**Device characterization.** All the device characterizations were performed in air under ambient conditions. Current-voltage (I-V) measurements were performed with a Keysight Semiconductor Parameter Analyzer (B1500A) with the devices kept in a shield box. The EQE was measured using a Newport Cornerstone 260 monochromator, a Thorlabs MC2000 chopper, a Stanford Research SR570 transimpedance amplifier and a Stanford Research SR830 lock-in amplifier. Calibrated Newport 818-UV and 818-IR photodetectors were used as the reference. For the noise measurements, devices were kept in a dark, shielded enclosure at room temperature under zero bias and directly connected with Stanford Research SR830 lock-in amplifier using its noise measurement function, following many other previous reports.[12,17,49,58–63] In order to measure the 3dB bandwidth, nanosecond laser (520 nm) was used as incident light, which was modulated by a waveform generator (Agilent 33220A) at various frequencies with 50% duty cycle. The output current was recorded with an oscilloscope. For linear dynamic range measurements, four-channel laser (Thorlabs) at 1310 nm was used as light source at a frequency of 7 Hz modulated by Agilent waveform generator. The light was directed by a beam-splitter to a Newport 818-IG detector and the device. The current was amplified by an SR570 amplifiers and recorded by an oscilloscope.

For image capture we used a Tungsten halogen lamp (1000W - 60° beam angle), placed at a distance of ~2m from the objects. The QD image sensor captured images of the objects at a distance of 1 – 1.5 m using a SWIR objective lens (VS-0814H1-SWIR) with a bandpass of 900 – 2000 nm. In addition to the sensitive range of the photodiode of 350nm to 1600nm, the imager captures the light from 350nm to 1600nm. The integration time was set at 10 ms and the applied revers bias of the pixels was 0.2 V.  Due to the Kapton tape induced edge effects, the images were cropped into 510×365 pixel resolution with ~100% effective pixel rate. Further optimization on etching process would lead to full VGA resolution.

**Data availability.**

All data that support the findings of this work are shown in the main text and Supplementary Information. Source data are provided with this paper.